# Interconnection network with a shared whiteboard: Impact of (a)synchronicity on computing power [*]

F. Becker[1], A. Kosowski[2], N. Nisse[3], I. Rapaport[4,5], and K. Suchan[6,7]

[1] LIFO, Universite d'Orléans, France
[2] LaBRI, INRIA Bordeaux Sud-Ouest, France.
[3] MASCOTTE, INRIA, I3S(CNRS/Univ. Nice Sophia Antipolis), France.
[4] Departamento de Ingeniería Matemática, Universidad de Chile, Santiago, Chile.
[5] Centro de Modelamiento Matemático (UMI 2807 CNRS), Universidad de Chile, Santiago, Chile.
[6] Facultad de Ingeniería y Ciencias, Universidad Adolfo Ibáñez, Santiago, Chile.
[7] Faculty of Applied Mathematics, AGH - University of Science and Technology, Cracow, Poland.

**Abstract.** In this work we study the computational power of graph-based models of distributed computing in which each node additionally has access to a global whiteboard. A node can read the contents of the whiteboard and, when activated, can write one message of $O(\log n)$ bits on it. When the protocol terminates, each node computes the output based on the final contents of the whiteboard. We consider several scheduling schemes for nodes, providing a strict ordering of their power in terms of the problems which can be solved with exactly one activation per node. The problems used to separate the models are related to Maximal Independent Set, detection of cycles of length 4, and BFS spanning tree constructions.

## 1 Introduction

A distributed system is a network where nodes correspond to agents or processors and links express the local knowledge of the nodes. To perform any calculation – like deciding some network's property – nodes may exchange information by interacting locally, i.e., with their neighbors. Since nodes lack of global knowledge, new algorithmic and complexity notions arise. In contrast with classical algorithmic theory – where the Turing machine is the consensus formal model of algorithm – in distributed systems many different models are considered. Under the paradigm that communication is much slower and more costly than the local computation, complexity analysis of distributed algorithms mainly focuses on message passing between the nodes. That is, an important performance measure is the number and the size of messages that are exchanged for performing some computation. Some theoretical models were conceived for studying particular aspects of protocols such as fault-tolerance, synchronism, locality, congestion, etc. One of the main questions arising is to determine the global properties of the network that can be computed locally.

In the model CONGEST [17], a network is represented by a graph whose nodes correspond to network processors and edges to inter-processor links. The communication is synchronous and occurs in discrete time rounds. In each round, each of the $n$ processors can send a message of size $O(\log n)$ bits through each of its outgoing links. A restriction of the CONGEST model has been proposed by Grumbach and Wu to study *frugal* computation [9]. In this model, where the total amount of information traversing each link is bounded by $O(\log n)$ bits, they showed that any first order logic formula can be evaluated in any planar or bounded degree network [9].

In [3], Becker *et al.* investigated a variation of CONGEST inspired by the Simultaneous Message Model defined by Babai, Kimmel and Lokam [2]. In this model, the total amount of local information that each node may provide is bounded by $O(\log n)$ bits. However, to compensate the little amount of knowledge that is shared, the communication is *global*: each node directly transmits its message to a central authority, *the referee*, that collects and uses them to answer some question about the network. This model allows to abstract away from the cost of *transmitting* data throughout the network, and to look at *how much local information must be shared* in order to compute some property.

---

[*] Partially supported by programs Fondap and Basal-CMM (I.R., K.S.), Fondecyt 1090156 (I.R.), Anillo ACT88 and Fondecyt 11090390 (K.S).

More precisely, each of the $n$ nodes, knowing only its own ID, the IDs of its neighbors and the size of the network, is allowed to send one message of $O(\log n)$ bits to the referee. Then, the referee can use the information it received to answer some question. Becker *et al.* asked whether this small amount of local information provided by each node is sufficient for the referee to decide some basic structural properties of the network topology $G$ [3]. For instance, simple questions like "Does $G$ contain a square?" or "Is the diameter of G at most 3?" cannot be solved. On the other hand, the referee can decode the messages in order to have full knowledge of $G$ when $G$ belongs to one of many graph classes such as planar graphs, bounded treewidth graphs and, more generally, bounded degeneracy graphs [3].

In this paper, we define natural extensions of the model in [3] and investigate the computational power of these new models.

**Computations using a shared whiteboard.** The computational model in [3] can be stated equivalently in the following form. Given a question on the topology of the network, every node writes *simultaneously* one message (computed from its local knowledge) on a global zone of shared memory, a *whiteboard*, and then, each of the nodes must answer the question by using only the content of the whiteboard.

In this paper, we intend to give more power to the initial model of [3]. For this purpose, we relax in different ways the simultaneity constraint. Roughly, messages may be written sequentially on the whiteboard. This allows nodes to compute their message taking into account the content of the whiteboard, i.e., the messages, or part of the messages, that have previously been written. In other words, in the new models we propose, nodes have extra ways to share information. Basically, the four models we now present aim at describing how the nodes can access the shared medium, in particular, differentiating synchronous and asynchronous networks.

The time is divided into discrete steps and, at every step, each node can see the current content of the whiteboard and may perform some local computation according to this content and its local knowledge. Along the evolution of the system, the nodes may be in three states: *awake*, *active*, or *terminated*. Initially, all nodes are *awake*. A node becoming *active* means that this node would like to write a message on the whiteboard, metaphorically speaking, it "rises its hand to speak". Finally, a node is in state *terminated* when its message has been written on the whiteboard. During one step, several awake nodes may become active and exactly one active node becomes terminated. To model the worst-case behavior of a model, the choice of the node that becomes terminated is done by an *adversary* among the set of active nodes. Note that a node may become active and terminated during one step. After the last step, when all nodes are terminated, all of them must be able to answer the question by using only the information stored on the whiteboard.

In this setting, we propose several scenarios leading to the definition of four computational models. A computational model is said *free* if, at any step, any awake node may decide to become active based on its knowledge and on its own protocol. On the other hand, the model is said *simultaneous* if all nodes are forced to become active during the first step. The other criterion we use to distinguish models is the state-transition during which a node must create the message it will eventually write on the whiteboard. In the *asynchronous* scenario, the nodes must create their message during the step when they become active. In the *synchronous* scenario, the nodes must create their message during the step when they become terminated.

Intuitively, in the asynchronous scenario, a node becoming active must compute its message regarding the content of the whiteboard at this step. However, this message is actually written on the whiteboard only when the node becomes terminated, which depends on the choice of the adversary. Thus, there may be some delay between the creation of a message and the step when it is written. In particular, the order in which the messages are created and the order in which they are actually available on the whiteboard may differ. In this way, we can model real-world asynchronous systems where there are no guarantees on the time of communications.

In this paper, we combine the free/simultaneous and asynchronous/synchronous scenarios and study the four resulting models. In particular, it is easy to check that the simultaneous-asynchronous model exactly corresponds to the model studied in [3]. On the other hand, the free-synchronous model is inspired by the Multiparty Communication Protocol introduced by Chandra, Furst and Lipton [4]. We aim at deciding which kind of problems can be solved in different models. Moreover, we intend to show that these models form a hierarchy in which the computation power increases strictly.

**Related work.** Many variations to the CONGEST model have been proposed in order to focus on different aspects of distributed computing. In a seminal paper, Linial introduced the LOCAL model [12, 17]. In the LOCAL model, the restriction on the size of messages is removed so that every vertex is allowed to send unbounded size messages in every round. This model focuses on the issue of locality in distributed systems, and more precisely on the question "*What cannot be computed locally?*" [11]. Difficult problems like minimum vertex cover and minimum dominating set cannot be well approximated when processors can locally exchange arbitrary long messages during a bounded number of rounds [11].

In the centralized computing point of view, testing a property $\mathcal{P}$ of a graph $G$ may consist of determining the minimum number of elementary queries (e.g., 'what is the $i^{th}$ neighbor of some vertex $v$?") necessary to decide whether the graph satisfies it. This model was first studied by Goldreich *et al.* [7] (see [6] for a survey). In this context, probabilistic algorithms are given that always accept a graph if it satisfies the property and reject with constant probability any graph that is "far enough" from the property. For instance, [1] gives lower and upper bounds for testing the triangle-freeness in general graphs. Closer to our model, Goldreich and Ron provide efficient algorithms for testing the connectivity of bounded degree graphs when the list of neighbors of every vertex is given [8].

Other trade-offs between the size of a data structure and the complexity (in terms of the number of bits that must be checked in this structure) of algorithms for solving some problems have been provided using the communication complexity model and the cell probe model [19, 20, 13, 14]. Testing graph properties has also been widely investigated using these frameworks (e.g., [18, 5, 16]).

## 2 Communication models

### 2.1 Protocol formulation

An interconnection network is modeled by a simple undirected connected $n$-node graph $G = (V, E)$. Each node $v \in V$ has a unique identifier $ID(v)$ between 1 and $n$. Typically, $V = \{v_1, \ldots, v_n\}$, where $v_i$ is such that $ID(v_i) = i$. Throughout the paper, a *graph* must be understood as a *labeled graph*.

At each node $v \in V$ there is an independent processing unit that knows the local knowledge of $v$: its own identifier, the identifier of each of its neighbors and the total number of nodes $n$. Moreover, any node $v \in V$ has a variable $output_v$ with arbitrary initial value and a variable $status_v \in \{awake, active, terminated\}$ initially set to $awake$.

Nodes can communicate with each other through a shared memory, called *whiteboard* and denoted by $\mathcal{B}$, that is initially empty. Any node can read $\mathcal{B}$ and any active node is allowed to write exactly one $O(\log n)$-bit message on it.

The time is divided into discrete steps. At any step, each node executes the same algorithm $\mathcal{A}$, or protocol, which may be divided into three sub-procedures. For any awake node $v$, the *activation* function takes the local knowledge of $v$ and the current content of $\mathcal{B}$ as input and decides to modify the $status_v$ variable to *active* or to maintain it as *awake*. For any active node $v$, the *message* function takes the local knowledge of $v$ and the current content of $\mathcal{B}$ as input and computes a message $m_v$ of $O(\log n)$ bits. In particular, we may assume that $m_v$ always contains the identifier of $v$ and the number of messages present on $\mathcal{B}$ at the step when $m_v$ is created. It can serve as a signature with the identity of the author and the "time" of its creation. When $n$ messages have been written on $\mathcal{B}$, the *decision* function takes the (final) content of $\mathcal{B}$ as input and computes the final value of $output_v$. While some nodes are not in the *terminated* state, an active node $v$ is chosen at the end of each time step and $m_v$ is written on $\mathcal{B}$, and then $status_v$ is set to *terminated*.

Given a problem[8], our goal is to design an algorithm $\mathcal{A}$ such that, at the end of the process (after the last step), and for any graph $G$, all nodes agree on the solution of the problem. That is, once all nodes have executed the decision function, all variables $output_v$, for any $v \in V$, give the correct answer for the problem. In what follows, to model the worst-case, the choice of which active node is chosen is done by an *adversary*. In this setting, we say that a problem *can be solved* in our model if there exists such an algorithm. It is important to note that, to avoid deadlock, a valid algorithm – precisely, its activation procedure – must ensure that, at any step, at least one node is active.

## 2.2 Scheduling of whiteboard access

We now specify four models of computation, following the general framework described above, that we study in this paper. For this purpose, we propose two additional constraints that may be satisfied or not, depending on the model.

In order to model how nodes access the whiteboard, i.e., in a synchronous way or not, we consider the following two variants: either a node must create its message as soon as it becomes active, or a node may defer creating its message until it has actually been chosen by the adversary to write it on the whiteboard. In the latter case, the node may take advantage of all the messages that have been written since it became active. More formally, we impose that any node executes the *message* function only once: in the *asynchronous* model, it must be executed during the same step when it becomes active, while in the *synchronous* model, it is executed when the node becomes terminated.

The second constraint we consider aims at avoiding an initial deadlock. Indeed, in some distributed contexts, it is not possible to decide whether it is better to write some message on the empty whiteboard or to wait a first message to be written. To deal with this, we consider that either the nodes may be free to decide when to become active, or they may be forced to be all active at the beginning. More formally, in the *simultaneous* model, the *activation* function is imposed to be the function that turns the status variable of each node to *active* during the first step (when the whiteboard is empty). In the *free* model, the definition of the *activation* function is part of the algorithm's design.

---

[8] Problems we are considering typically consist of a question on the graph's topology and the output of which may be a boolean value (decision problems), the adjacency matrix of the graph, etc. However, our models may also be used to deal with classical distributed problems like consensus, leader election, etc.

|  | message created when node becomes active | message created when node is chosen |
|---|---|---|
| all nodes initially active | SimAsync | SimSync |
| no node initially active | FreeAsync | FreeSync |

**Table 1.** Classification of communication models.

To satisfy or not this pair of constraints leads to four different communication models. We now detail the four communication models studied in this work. These models are summarized in Table 1.

*Simultaneous Asynchronous* (SimAsync). In this model, all nodes become active and create their messages at the first step. In other words, all nodes create their messages while the whiteboard is still empty. Hence, the message created by a node $v$ only depends on the local knowledge of $v$ and on $n$. Moreover, the ordering in which the messages are written on $\mathcal{B}$ (i.e., in which the nodes are chosen by the adversary) is clearly not relevant. This is exactly the communication model studied in [3].

*Simultaneous Synchronous* (SimSync). In this model, all nodes become active in the first step. However, messages are created just before being written on the whiteboard. In other words, the adversary chooses an ordering $(x_1, \ldots, x_n)$ of $V$ such that, at step $1 \leq i \leq n$, vertex $x_i$ computes its message according to its local knowledge, the total number of nodes $n$, and the $i-1$ messages that have previously been written by $x_1, \ldots, x_{i-1}$ on $\mathcal{B}$. Note that the ordering is not known *a priori* by the nodes. Hence, an algorithm for solving a problem in this model must solve it for any ordering chosen by the adversary.

*Free Asynchronous* (FreeAsync). In this model, the nodes decide when to turn active and create their messages at the same step that they become active. That is, a node may create a message long before being chosen by the adversary to write it down on $\mathcal{B}$.

*Free Synchronous* (FreeSync). The nodes decide when to become active and create their messages only when they are asked by the adversary to write them on $\mathcal{B}$.

This paper aims at deciding what kind of problems can be solved in each of these models. For instance, [3] proves that deciding if a graph has degeneracy $k$, $k \geq 1$, can be solved in SimAsync. On the opposite, deciding whether a graph contains a triangle as a subgraph and deciding whether a graph has diameter at most 3 cannot be solved in SimAsync [3].

In Section 3, we focus on problems that separate the models, i.e., that can be solved in some model but not in another one. In Section 4, we focus on problems related to connectivity.

First of all, we prove the following lemma that extends a result of [3]. Let BUILD be the problem that consists in computing the adjacency matrix of an input graph $G$.

**Lemma 1.** *Let $\mathcal{G}$ be a family of $n$-node labeled graphs, and $g(n)$ be the number of graphs in $\mathcal{G}$. In any of the four considered models, BUILD can be solved in the class $\mathcal{G}$ only if $\log g(n) = O(n \log n)$.*

*Proof.* Consider any algorithm in one of the four considered models. In any model, at the end of the communication process, $n$ messages of size $O(\log n)$ bits are written on $\mathcal{B}$. Notice that includes the information on the order in which the messages have been created. Hence, at the end, a total of $O(n \log n)$ bits are available on the whiteboard. For any node to distinguish two different graphs in $\mathcal{G}$, we must have $\log g(n) = O(n \log n)$. □

For the ease of descriptions, in what follows we will not define explicitly the functions for activation, message creation and decision. Nevertheless, they always will be clear from the context.

## 3 A strict hierarchy

In this section, we intend to show that these models form a hierarchy in which the computation power strictly increases. Let us formalize this idea. We say that model $Y$ is *more powerful* than model $X$, denoted by $X \leq Y$, if every problem that can be solved in $X$ can also be solved in $Y$. Moreover, $Y$ is *strictly more powerful* than $X$, denoted by $X < Y$, if $X \leq Y$ and there exists a problem $\mathcal{P}$ that can be solved in $Y$ but not in $X$.

The main result of this section is the following theorem:

**Theorem 1.** SimAsync < SimSync < FreeAsync $\leq$ FreeSync.

We start with the following weaker result:

**Proposition 1.** SimAsync $\leq$ SimSync $\leq$ FreeAsync $\leq$ FreeSync.

*Proof.* Given two models $X, Y$, to show that $X \leq Y$, we consider an algorithm for solving some problem in $X$ and show how to turn this algorithm to satisfy requirements of $Y$.

- SimAsync $\leq$ SimSync. It suffices that nodes in the SimSync protocol ignore the messages present on the whiteboard when they create their own.
- SimSync $\leq$ FreeAsync. Recall that a problem is solved in the SimSync model if the nodes compute the output *no matter* the order chosen by the adversary. So we can translate a SimSync protocol into a FreeAsync one if we fix an order (for instance $v_1, \ldots, v_n$) and use this order for a sequential activation of the nodes.
- FreeAsync $\leq$ FreeSync. It is the situation of the first inequality. It suffices to force the protocols in FreeSync to create their messages based only on what was known at the moment when they became active.

□

### 3.1 SimAsync vs. SimSync

We consider here a "rooted" version of the Inclusion Maximal Independent Set problem. This problem, denoted by MIS, takes as input an $n$-node graph $G = (V, E)$ together with an identifier $ID(x)$, $x \in V$, and the desired output is any maximal (by inclusion) independent set containing $x$.

**Proposition 2.** MIS *can be solved in the* SimSync *model.*

*Proof.* Recall that in the adversarial model, all nodes are initially active and that the adversary chooses the ordering in which the nodes write their messages. Hence, an algorithm in this model must specify the message created by a node $v$, according to the local knowledge of $v$ and the messages written on the whiteboard before $v$ is chosen by the adversary.

The protocol is trivial (it is the greedy one). When node $v$ is chosen by the adversary, the message of $v$ is either its own ID (meaning that $v$ belongs to the final independent set) or $v$ writes "no" (otherwise). The choice of the message is done as follows. The message is $ID(v)$ either if $v = x$ or if $v \notin N(x)$ and $ID(y)$ does not appear on the whiteboard for any $y \in N(v)$. Otherwise, the message of $v$ is "no".

Clearly, at the end, the set of vertices with their IDs on the whiteboard consists of an inclusion maximal independent set containing $x$. □

**Proposition 3.** MIS *cannot be solved in the* SIMASYNC *model.*

*Proof.* We proceed by contradiction. Let us assume that there exists a protocol $\mathcal{A}$ for solving MIS in the SIMASYNC model. Then we show how to design an algorithm $\mathcal{A}'$ to solve the BUILD Problem for any graph in this model, contradicting Lemma 1.

Let $G = (V, E)$ be a graph with $V = \{v_1, \ldots, v_n\}$. For any $1 \leq i < j \leq n$, let $G_{i,j}^{(x)}$ be obtained from $G$ by adding a vertex $x$ adjacent to every vertex in $V$ with the exception of $v_i$ and $v_j$. Note that $\{x, v_i, v_j\}$ is the only inclusion maximal independent set containing $x$ in $G_{i,j}^{(x)}$ if and only if $\{v_i, v_j\} \notin E$. Indeed, if $\{v_i, v_j\} \in E$, there are two inclusion maximal independent sets containing $x$: $\{x, v_i\}$ and $\{x, v_j\}$.

Recall that, in the SIMASYNC model, all nodes must create their message initially, i.e., while the whiteboard is still empty. Hence, the message created by a node only depends on its local knowledge. We denote by $\mathcal{A}(v_k, G_{i,j}^{(x)})$ the message created by node $v_k$ following protocol $\mathcal{A}$ when the input graph is $G_{i,j}^{(x)}$.

Notice that, for a given $k$, the node $v_k$ can generate only two possible messages $\mathcal{A}(v_k, G_{i,j}^{(x)})$ depending on whether $k \in \{i, j\}$ or $k \notin \{i, j\}$. Therefore, we call $m_k$ the message that $v_k$ generates when $k \in \{i, j\}$ (i.e., $x$ and $v_k$ are not neighbors) and $m'_k$ the message $v_k$ generates when $k \notin \{i, j\}$ (i.e., $x$ and $v_k$ are neighbors).

From the previous protocol $\mathcal{A}$ we are going to define another protocol $\mathcal{A}'$ in the SIMASYNC model which solves the BUILD Problem for any graph. Protocol $\mathcal{A}'$ works as follows. Every node $v_k$ generates the pair $(m_k, m'_k)$ of the two messages $v_k$ would send in $\mathcal{A}$ when it is adjacent to $x$ and when it is not. Clearly, this consists of $O(\log n)$ bits.

Now let us prove that any node can reconstruct $G = (V, E)$ from the messages generated by $\mathcal{A}'$. More precisely, for any $1 \leq s < t \leq n$, any node can decide whether $\{v_s, v_t\} \in E$ or not. It is enough for any node to simulate de decision function of $\mathcal{A}$ in $G_{s,t}^{(x)}$ by using messages $m_s, m_t$ and $\{m'_k \ : \ k \in \{1, \cdots, n\} \setminus \{s, t\}\}$. Since the output of $\mathcal{A}$ is $\{x, v_s, v_t\}$ if and only if $\{v_s, v_t\} \notin E$, the results follows.

This means that from $O(n \log n)$ bits we can solve BUILD in the class of all graphs - a contradiction. □

**Corollary 1.** SIMASYNC < SIMSYNC.

We discuss now another problem that could possibly separate the two models. Given an $(n-1)$-regular $2n$-node graph $G$, the 2-CLIQUES problem consists in deciding whether $G$ is the disjoint union of two complete graphs with $n$ vertices or not.

It is easy to show that 2-CLIQUES can be solved in the SIMSYNC model. Indeed, a trivial protocol can partition the vertices into two cliques numbered 0 and 1 if the input consists of two cliques, or otherwise indicate that it is not the case. The first vertex $f$ to be chosen by the adversary writes $(ID(f), 0)$ on $\mathcal{B}$. Then, each time a vertex $v$ is chosen, it writes $(ID(v), 0)$ if it "believes" to be in the same clique as $f$, and $(ID(v), 1)$ otherwise. More precisely, let $S_v$ be the subset of neighbors of $v$ that have already written a message on the whiteboard. If $S_v = \emptyset$ then $v$ writes 1. If all nodes in $S_v$ have written that they belong to the the same clique $c \in \{0, 1\}$ then $v$ writes $c$, and $v$ writes "no" otherwise. Clearly, $G$ is the disjoint union of two cliques if and only if there is no message "no" on the whiteboard at the end of the communication process.

Proving that 2-CLIQUES cannot be solved in the SIMASYNC model is an interesting question because it would allow us to show that CONNECTIVITY (deciding whether a graph is connected or not) cannot be solved in the SIMASYNC model. Indeed, it is easy to show that an $(n-1)$-regular $2n$-node graph is the disjoint union of two cliques if and only if it is not connected. We leave this as an open question:

**Open Problem 1.** Can 2-CLIQUES be solved in the SIMASYNC model?

## 3.2 SimSync vs. FreeAsync

Let SQUARE be the problem that consists in deciding whether a graph $G$ contains a *square* (induced or not), i.e., whether $V(G)$ contains four vertices $a, b, c$ and $d$ such that $a$ is adjacent to $b$ which is adjacent to $c$ which is adjacent to $d$ which is adjacent to $a$.

In [3], it is proved that SQUARE cannot be solved in the SIMASYNC model. The proof consists of showing that if SQUARE could be solved in the SIMASYNC model, then BUILD could be solved, in this model, in the class of square-free graphs, a contradiction.

**Theorem 2.** *[3]* SQUARE *cannot be solved in the* SIMASYNC *model.*

We extends this result to the SIMSYNC model. More precisely, we show that SQUARE cannot be solved in SIMSYNC model even restricted to a specific class of graphs. We then show that, in this particular graph class, SQUARE can be solved in the FREEASYNC model. Hence, the SQUARE problem on this class of graphs separates SIMSYNC and FREEASYNC.

First, let $\mathcal{C}$ be the class of graphs of even order $N = 2n$ ($n \geq 1$) that can be obtained as follows. $G \in \mathcal{C}$ has the vertex set $\{v_1, \cdots, v_n, v_{n+1}, \cdots, v_{2n}\}$ where $G[\{v_1, \cdots, v_n\}]$ induces a square-free $n$-node graph $H$, and, for any $1 \leq i \leq n$, $v_{i+n}$ is adjacent to $v_i$, and finally, there is a unique additional edge between $v_{n+i}$ and $v_{n+j}$ for some $1 \leq i < j \leq n$ (i.e., for any $k \in \{1, \cdots, n\} \setminus \{i, j\}$, $v_{n+k}$ has degree one in $G$). In the following, we note $G = (H, i, j)$. Note that, since there are $\Omega(2^{n^{3/2}})$ (labeled) square-free graphs with $n$ nodes [10], $|\mathcal{C}| = \Omega(2^{n^{3/2}}) = \Omega(2^{N^{3/2}})$.

**Proposition 4.** SQUARE *cannot be solved in the* SIMSYNC *model, even when inputs are restricted to* $\mathcal{C}$.

*Proof.* For purpose of contradiction, let us assume that there is a protocol $\mathcal{P}$ for solving SQUARE in $\mathcal{C}$ in the SIMSYNC model. We design a protocol $\mathcal{P}'$ for solving BUILD in $\mathcal{C}$, contradicting Lemma 1 since $N \log N = o(\log |\mathcal{C}|)$.

In the SIMSYNC model, the nodes are asked to write their messages in some order. When it is the turn of node $v_i$ it computes its message according to the current content of the whiteboard and to its own neighborhood.

Let $G = (H, \ell, k) \in \mathcal{C}$. The key point is that $\mathcal{P}$ solves SQUARE in $G$ whatever be the order in which the vertices write their messages. In particular, if the vertices in $\{v_1, \cdots, v_n\}$ are interrogated first, their messages cannot bring any information on the single edge between two vertices in $\{v_{n+1}, \cdots, v_{2n}\}$. Moreover, the vertices in $\{v_{n+1}, \cdots, v_{2n}\}$ have degree at most two in $G$ and so can write their full neighborhood. Using these two facts, we design $\mathcal{P}'$ that will be used to rebuild any graph in $\mathcal{C}$.

The protocol $\mathcal{P}'$ is defined as follows. For any $i \leq 2n$, the message created and written by node $v_i$ when it is interrogated consists of

- if $n < i \leq 2n$, then $v_i$ writes the identifiers of its at most two neighbors;
- otherwise, let $\mathcal{O}$ be the sequence of the vertices in $\{v_1, \cdots, v_n\}$ that have been interrogated before $v_i$, in order. Using its local neighborhood and the messages previously written by the vertices in $\mathcal{O}$, $v_i$ writes the message it would have written following $\mathcal{P}$ and in the ordering $\mathcal{O}$.

We now show that whatever be the ordering $\mathcal{O}$ in which the $2n$ vertices have been interrogated, the final content of the whiteboard allows any vertex to build the adjacency matrix of $G$. More precisely, we show that any node can decide whether the edge $\{v_i, v_j\} \in E(G)$ for any $1 \leq i < j \leq 2n$.

Clearly, for any $1 \leq i \leq n$, the edges adjacent to $v_{n+i}$ can be decided since they appear explicitly on the whiteboard (in particular for $i = \ell$ and $j = k$). Let $\mathcal{O}'$ be the restriction of $\mathcal{O}$ to the vertices in $\{v_1, \cdots, v_n\}$. Now, for any $1 \leq i < j \leq n$, we show that, using the information on the whiteboard, any node can simulate the protocol $\mathcal{P}$ in the graph $(H, i, j)$, against the ordering $\mathcal{O}' \odot (v_{n+1}, \cdots, v_{2n})$ and decide whether $\{v_i, v_j\}$ does exist in $G$ which is the case if and only if $(H, i, j)$ has a square.

After the execution of the protocol, the whiteboard contains the messages $m_1, \cdots, m_n$ that would have written by $v_1, \cdots, v_n$ when following protocol $\mathcal{P}$ against the ordering $\mathcal{O}'$. Using this information, any node can compute the message that $v_t$ ($n < t \leq 2n$) would have written when executing $\mathcal{P}$ in $(H, i, j)$. Therefore, it can decide whether $(H, i, j)$ has a square, i.e., whether $\{v_i, v_j\} \in E(G)$.

Since by Lemma 1 and because of the cardinality of $\mathcal{C}$, no protocol can solve BUILD in $\mathcal{C}$ and in the SIMSYNC model, we get a contradiction. □

**Corollary 2.** SQUARE *cannot be solved in the* SIMSYNC *model.*

**Proposition 5.** SQUARE *can be solved in* $\mathcal{C}$ *in the* FREEASYNC *model.*

*Proof.* The protocol is almost trivial. First, any node with identifiant at least $n + 1$ becomes active and creates a message containing its neighborhood (recall that it has at most two neighbors in $G$). By reading the $n$ messages that have been written by $v_{n+1}, \cdots, v_{2n}$, all remaining nodes know the (unique) pair $(i, j)$ such that $\{v_{i+n}, v_{j+n}\} \in E(G)$ and know the neighbor $u$ of $v_{n+i}$, resp., the neighbor $v$ of $v_{j+n}$, in $\{v_1, \cdots, v_n\}$. Finally, all remaining vertices become active: for any $k \leq n$, $v_k$ writes an empty message if $v_k \notin \{u, v\}$ and $u$ and $v$ write "yes" if they are adjacent and "no" otherwise. Clearly, the graph admits a square if and only if the whiteboard eventually contains "yes". □

**Corollary 3.** SIMSYNC<FREEASYNC.

**Open Problem 2.** Can SQUARE be solved in the FREEASYNC model, in FREESYNC?

## 4 Connectivity and related problems

One of the main questions arising in distributed environment concerns connectivity. For instance, one important task in wireless network consists in computing a connected spanning subgraph (e.g., a spanning tree) the links of which will be used for communications. In this section, we ask in which of our models such problems can be solved.

We consider the following three problems. By increasing level of difficulty[9]: the CONNECTIVITY Problem asks if an input graph is connected or not; given an input graph $G$ and an input identifier $ID(r)$, the SPANNING-TREE Problem requires as output a spanning-tree of $G$ rooted in $r$ if it exists; similarly, the BFS Problem requires a BFS-tree of the input graph $G$ rooted in some given node (if $G$ is connected) as an output.

In [3] the authors conjecture that CONNECTIVITY (and therefore SPANNING-TREE) cannot be solved in the SIMASYNC model. This is still an open question. We do not even know whether these problems can be solved in the SIMSYNC model.

Nevertheless, it is clear that SPANNING-TREE (and therefore CONNECTIVITY) can be solved in the FREEASYNC model. The protocol is the greedy one. First the root $r$ becomes active and is chosen by the adversary and writes its ID. Then, at each step, nodes having a neighbor already included in the spanning tree become active and choose its parent.

---

[9] We say that a problem $\mathcal{P}$ is more difficult than a problem $\mathcal{P}'$ if any algorithm for solving $\mathcal{P}$ also solves $\mathcal{P}'$, or equivalently, any solution for $\mathcal{P}$ is a solution for $\mathcal{P}'$

In the more powerful FreeSync model it is possible not only to construct a spanning tree but a BFS tree. Table 2 summarizes the results of this section.

|               | SimAsync | SimSync | FreeAsync                      | FreeSync |
|---------------|----------|---------|--------------------------------|----------|
| BFS           | No       | No      | ?                              | Ok       |
|               |          | (Prop. 7) | Ok in Bipartite graphs (Cor. 4) | (Th. 6)  |
| Spanning-tree | ?        | ?       | Ok (Remark above)              | Ok       |
| Connectivity  | ?        | ?       | Ok                             | Ok       |

**Table 2.** Various problems related to connectivity and models where they can(not) be solved.

**Proposition 6.** BFS *can be solved in the* FreeSync *model.*

*Proof.* The idea is to simulate phases at the end of which every node in the same layer, i.e., at the same distance from $v_1$, knows the ID of its parent. The protocol must let the nodes of layer $k$, i.e., at distance $k \geq 1$ from $v_1$, to know the step when *all the nodes of layer $k-1$* have already written their messages on the whiteboard.

The algorithm for solving BFS is defined as follows. Initially (when the whiteboard is empty), only $v_1$ must become active and writes its ID, its degree and its layer 0. This is Phase 0.

At Phase $i > 0$, any node $v$ in layer $i$ becomes active and its message consists of

1. its own ID;
2. its layer $i$;
3. the ID of its neighbor in layer $i-1$ with minimum ID (this neighbor will be considered as its parent in the BFS tree);
4. the number $a_v$ of its neighbors in layer $i-1$;
5. the number $b_v$ of its neighbors which are not in layer $i-1$ ($a_v + b_v$ equals the degree of $v$);
6. the number $c_v$ of its neighbors in layer $i$ that have already written their messages on the whiteboard.

We now have to prove that every node becomes active at the right phase and that it can compute the required information from its local knowledge and what have previously been written on the whiteboard. We prove it by induction on $i \geq 1$. In particular, we prove that all nodes can decide when Phase $i-1$ terminates and that, at this step, *the number $e_{i-1}$ of edges between layer $i-1$ and layer $i$* can be computed from the information available on the whiteboard.

Phase 0 terminates when $v_1$ writes its message which contains its degree, i.e., $e_0$. Then, all neighbors of $v_1$ (the vertices of layer 1) become active. The vertices of $N(v_1)$ can easily compute the required information, no matter the ordering the adversary chooses. Note that information 6 can be obtained because we consider the FreeSync model, i.e., any node can create its message at the step when it is chosen (and not when it becomes active). Moreover, all nodes know the number of vertices in layer 1 and so can decide when Phase 1 terminates. Finally, $e_1$ is exactly the sum of $b_v - 2c_v$ among the vertices $v$ in layer 1. Hence, at the end of Phase 1, the induction hypothesis is satisfied.

Assume the induction hypothesis is satisfied at the end of Phase $i > 0$. Then, any node that has not become active yet and that has a neighbor in layer $i$ knows that it belongs to layer $i+1$ and becomes active. No matter the ordering the adversary chooses, the vertices in layer $i+1$ can easily compute the required information since, by the induction hypothesis, all nodes in layer $i$ have written their IDs together with the corresponding layer on the whiteboard.

Moreover, the nodes can detect the end of Phase $i+1$ since $e_i$ exactly equals the sum of the $a_v$ over the vertices $v$ in layer $i+1$. Finally, $e_{i+1}$ is exactly the sum of $b_v - 2c_v$ among the vertices $v$ in layer $i+1$.

To conclude, since the vertices know $n$, they can detect when the communication process terminates. Moreover, any node can compute a BFS tree because every vertex has written its parent ID on the whiteboard. □

**Corollary 4.** BFS *can be solved in the* FREEASYNC *model, in the class of bipartite graphs.*

*Proof.* In a bipartite graph there are no edges between nodes in the same layer, and therefore $c_v = 0$ for every node $v$. In other words, we need to apply the protocol for the general case without computing information 6. □

**Proposition 7.** BFS *cannot be solved in the* SIMSYNC *model.*

*Proof.* For purpose of contradiction, let us assume that there is a protocol $\mathcal{P}$ for solving BFS in the SIMSYNC model. We design a protocol $\mathcal{P}'$ for solving BUILD for any $n$-node graph, contradicting Lemma 1.

First, let $\mathcal{C}$ be the class of graphs of even order $N = 4n - 1$ ($n \geq 1$) that can be obtained as follows. A graph $G = (H, i) \in \mathcal{C}$ is built from any $n$-node graph $H$ with vertex-set $\{v_1, \cdots, v_n\}$ and any integer $i$, $1 \leq i \leq n$, in the following way: let us add $3n$ vertices $\{v_{n+1}, \cdots, v_{4n}\} = \{r = v_{n+1}, a_1, \cdots, a_n, b_1, \cdots, b_{i-1}, b_{i+1}, \cdots, b_n, c_1, \cdots, c_{i-1}, c_{i+1}, \cdots, c_n\}$ such that, for any $j \leq n$, $r$ is adjacent to $a_j$, for any $j \leq n$, $j \neq i$, $b_j$ is adjacent to $v_j$, $c_j$ is adjacent to $a_j$ and to $b_j$, and finally, $a_i$ is adjacent to $v_i$. Note that $|\mathcal{C}| = \Omega(2^{n^2})$.

Let $G = (H, k) \in \mathcal{C}$. The key point is that $\mathcal{P}$ solves BFS in $G$ whatever be the order in which the vertices write their messages. In particular, if the vertices in $\{v_1, \cdots, v_n\}$ are interrogated first, their messages cannot bring any information on which of the $a_i$ is adjacent to $r$. The neighborhood of $r$ can easily be encoded with $O(\log n)$ bits since it is adjacent to all $a_i$, $i \leq n$. Moreover, the vertices in $\{v_{n+2}, \cdots, v_{4n}\}$ have degree two in $G$ and so can write their full neighborhood. Using these two facts, we design $\mathcal{P}'$ that will be used to rebuild any graph in $\mathcal{C}$.

The protocol $\mathcal{P}'$ is defined as follows. For any $i \leq 4n$, the message created and written by node $v_i$ when it is interrogated consists of

- if $i = n+1$, $r$ writes its neighborhood.
- if $n+1 < i \leq 4n$, then $v_i$ writes the identifiers of its at most two neighbors;
- otherwise, let $\mathcal{O}$ be the sequence of the vertices in $\{v_1, \cdots, v_n\}$ that have been interrogated before $v_i$, in order. Using its local neighborhood and the messages previously written by the vertices in $\mathcal{O}$, $v_i$ writes the message it would have written following $\mathcal{P}$ and in the ordering $\mathcal{O}$.

We now show that whatever be the ordering $\mathcal{O}$ in which the $4n$ vertices have been interrogated, the final content of the whiteboard allows any vertex to build the adjacency matrix of $G$. More precisely, we show that any node can decide whether the edge $\{v_i, v_j\} \in E(G)$ for any $1 \leq i < j \leq 2n$.

Clearly, for any $1 \leq i \leq 3n$, the edges adjacent to $v_{n+i}$ can be decided since they appear explicitly on the whiteboard. Let $\mathcal{O}'$ be the restriction of $\mathcal{O}$ to the vertices in $\{v_1, \cdots, v_n\}$. Now, for any $1 \leq i < j \leq n$, we show that, using the information on the whiteboard, any node can simulate the protocol $\mathcal{P}$ in the graph $(H, i)$, against the ordering $\mathcal{O}' \odot (v_{n+1}, \cdots, v_{4n})$. By construction of $(H, i)$, if the edge $\{v_i, v_j\} \in E(G)$ it must belong to the BFS-tree computed by $\mathcal{P}$ on $(H, i)$. Hence, any node can decide whether $\{v_i, v_j\}$ does exist in $G$.

More precisely, after the execution of the protocol, the whiteboard contains the messages $m_1, \cdots, m_n$ that would have written by $v_1, \cdots, v_n$ when following protocol $\mathcal{P}$ against the ordering $\mathcal{O}'$. Using this information, any node can compute the message that $v_t$ ($n < t \leq 4n$) would have written when executing $\mathcal{P}$ in $(H, i)$. Therefore, it can decide whether $\{v_i, v_j\}$ belongs to the BFS-tree of $(H, i)$, i.e., whether $\{v_i, v_j\} \in E(G)$.

Since by Lemma 1 and because of the cardinality of $\mathcal{C}$, no protocol can solve BUILD in $\mathcal{C}$ and in the SIMSYNC model, we get a contradiction. □

**Open Problem 3.** Is it true that FREEASYNC < FREESYNC? We conjecture that this is the case and that in fact BFS cannot be solved in the FREEASYNC model.

## 5 Conclusion and perspectives

We have investigated four models of distributed computing, extending the results presented in [3]. The definitions of these models are based on some intuitive conditions related to synchronicity, and we have seen that imposing them on the system has significant impact on what problems can be solved. We proved that there exists a hierarchy of non-decreasing computing power between these models. Moreover, we proved that in two cases the power strictly increases and left an open problem to check if it also holds in the third case.

We have analyzed several problems related to independence, cyclicity and connectivity. For these problems, we ask what are the conditions that a distributed computational model requires to solve them. Motivated by applications in routing [15], we payed special attention to connectivity: we analyzed the complexity of general spanning tree and BFS-tree construction. In particular, we showed that BFS-tree construction cannot be solved if nodes are not free to decide when they activate. On the other hand, the problem can be solved under the additional condition of synchronicity. The necessity of synchronicity for BFS-tree construction is left as an open problem.

We only analyzed the systems where no faults are allowed. It would be interesting to see what can be done when the system admits computation or communication errors. Another possible direction is to further the analysis of construction problems for connected spanning subgraphs that satisfy properties needed in compact routing protocols (see [15]).

## References

1. Noga Alon, Tali Kaufman, Michael Krivelevich, and Dana Ron. Testing triangle-freeness in general graphs. *SIAM J. Discrete Math.*, 22(2):786–819, 2008.
2. László Babai, Anna Gál, Peter G. Kimmel, and Satyanarayana V. Lokam. Communication complexity of simultaneous messages. *SIAM J. Comput.*, 33:137–166, 2004.
3. Florent Becker, Martin Matamala, Nicolas Nisse, Ivan Rapaport, Karol Suchan, and Ioan Todinca. Adding a referee to an interconnection network: What can(not) be computed in one round. In *Parallel and Distributed Processing Symposium, International*, pages 508–514. IEEE Computer Society, 2011.
4. Ashok K. Chandra, Merrick L. Furst, and Richard J. Lipton. Multi-party protocols. In *Proceedings of the fifteenth annual ACM symposium on Theory of computing*, STOC '83, pages 94–99. ACM, 1983.
5. Yevgeniy Dodis and Sanjeev Khanna. Space time tradeoffs for graph properties. In *Proceedings of the 26th International Colloquium on Automata, Languages and Programming (ICALP)*, volume 1644 of *Lecture Notes in Computer Science*, pages 291–300. Springer, 1999.
6. Eldar Fischer. The art of uninformed decisions: A primer to property testing. *Bulletin of the EATCS*, 75:97–126, 2001.
7. Oded Goldreich, Shafi Goldwasser, and Dana Ron. Property testing and its connection to learning and approximation. *J. ACM*, 45(4):653–750, 1998.
8. Oded Goldreich and Dana Ron. Property testing in bounded degree graphs. *Algorithmica*, 32(2):302–343, 2002.

9. Stéphane Grumbach and Zhilin Wu. Logical locality entails frugal distributed computation over graphs (extended abstract). In *Proceedings of 35th International Workshop on Graph-Theoretic Concepts in Computer Science (WG)*, volume 5911 of *Lecture Notes in Computer Science*, pages 154–165, 2009.
10. Daniel J. Kleitman and Kenneth J. Winston. On the number of graphs without 4-cycles. *Discrete Mathematics*, 41(2):167–172, 1982.
11. Fabian Kuhn, Thomas Moscibroda, and Roger Wattenhofer. What cannot be computed locally! In *Proceedings of the 23rd Annual ACM Symposium on Principles of Distributed Computing (PODC)*, pages 300–309. ACM, 2004.
12. Nathan Linial. Locality in distributed graph algorithms. *SIAM J. Comput.*, 21(1):193–201, 1992.
13. Peter Bro Miltersen. The bit probe complexity measure revisited. In *Proceedings of the 10th Annual Symposium on Theoretical Aspects of Computer Science (STACS)*, volume 665 of *Lecture Notes in Computer Science*, pages 662–671. Springer, 1993.
14. Peter Bro Miltersen, Noam Nisan, Shmuel Safra, and Avi Wigderson. On data structures and asymmetric communication complexity. In *Proceedings of the 27th Annual ACM Symposium on Theory of Computing (STOC)*, pages 103–111. ACM, 1995.
15. Nicolas Nisse, Ivan Rapaport, and Karol Suchan. Distributed computing of efficient routing schemes in generalized chordal graphs. In *SIROCCO*, volume 5869 of *Lecture Notes in Computer Science*, pages 252–265. Springer, 2009.
16. Mihai Patrascu and Erik D. Demaine. Logarithmic lower bounds in the cell-probe model. *SIAM J. Comput.*, 35(4):932–963, 2006.
17. David Peleg. *Distributed computing: a locality-sensitive approach*. SIAM Monographs on Discrete Mathematics and Applications, 2000.
18. Prasoon Tiwari. Lower bounds on communication complexity in distributed computer networks. *J. ACM*, 34(4):921–938, 1987.
19. Andrew Chi-Chih Yao. Some complexity questions related to distributive computing (preliminary report). In *Proceedings of the 11th Annual ACM Symposium on Theory of Computing (STOC)*, pages 209–213. ACM, 1979.
20. Andrew Chi-Chih Yao. Should tables be sorted? *J. ACM*, 28(3):615–628, 1981.